\documentclass[12pt]{article}
\usepackage{graphicx}\usepackage{amssymb}\usepackage{bm}\usepackage{amsfonts}\usepackage{amsmath}\usepackage{cases}\usepackage{hyperref}
\newcommand{\beq}{\begin{equation}}\newcommand{\eeq}{\end{equation}}\newcommand{\beqa}{\begin{eqnarray}}
\newcommand{\eeqa}{\end{eqnarray}}\newcommand{\del}{\nabla}

\newcommand{\p}{\partial}
\newcommand{\G}{\Gamma}
\newcommand{\Del}{\nabla}

\begin{document}
{\renewcommand{\thefootnote}{\fnsymbol{footnote}}

\begin{center}
{\LARGE  Is geometry bosonic or fermionic?}\\
\vspace{1.5em}
Taylor L. Hughes\footnote{e-mail address: {\tt hughest@illinois.edu}} and Andrew Randono\footnote{e-mail address: {\tt arandono@perimeterinstitute.ca} }
\\
\vspace{0.5em}
$^{\ast}$Department of Physics at the University of Illinois at Urbana Champaign\\
1110 West Green St. \\
Urbana, IL 61801, USA\\
\vspace{0.5em}
$^{\dagger}$The Perimeter Institute for Theoretical Physics \\
31 Caroline Street North\\
Waterloo, ON N2L 2Y5, Canada
\vspace{1.5em}
\end{center}
}

\setcounter{footnote}{0}

\begin{abstract}
It is generally assumed that the gravitational field is bosonic. Here we show that a simple propagating torsional theory can give rise to localized geometric structures that can consistently be quantized as fermions under exchange. To demonstrate this, we show that the model can be formally mapped onto the Skyrme model of baryons, and we use well-known results from Skyrme theory. This begs the question: {\it Is geometry bosonic or fermionic (or both)?}

\end{abstract}

\section*{Introduction} It is generally taken on faith that geometry is described by bosonic fields on a manifold. In the context of quantum gravity this presumption is usually buttressed by the simple observation that perturbative quantization of gravity describes interacting spin-2 gravitons. Being spin-2, the standard application of the spin-statistics theorem teaches us that the graviton should be bosonic. And so should be, it is presumed, any complicated geometric structures that are built from bosonic gravitons. 

On the other hand, it is well known (at least in some circles) that a sufficiently non-linear system can give rise to emergent degrees of freedom where the association of exchange statistics to particle-like field configurations is considerably more subtle. As these new degrees of freedom are generally of a topological nature, the true statistics of the quantized system can only be determined non-perturbatively. It can happen, for example, that systems whose perturbative excitations involve strictly bosonic fields, also possess emergent excitations that are fermionic in nature.

Therefore it is possible that a dynamical theory of geometry contains both bosonic and fermionic objects\footnote{Or potentially even more exotic statistics.}. In this paper we take {\it geometry} to include torsion in addition to the ordinary Riemannian geometry of a metric field. We will focus on a particular, well-motivated, model of dynamical torsion in a regime where the Riemannian curvature is negligible thus allowing us to isolate the torsional degrees of freedom. As we will show, the non-linear structure of the theory can give rise to both bosonic and fermionic quanta. Furthermore, the topological modes of the system have the peculiar feature that they mimic well-known particulate structures of the Standard Model, and as such the model could provide a fresh approach to the {\it matter-from-geometry} paradigm \cite{Arnsdorf:1998vq,AsselmeyerMaluga:2010zh,BilsonThompson:2006yc,Friedman:1980st,Friedman:1982du,Giulini:2009ts}. 

\subsection*{Motivation}
{\it Why should we consider a propagating torsion theory, and how can we draw conclusions from it about geometry in general?} 

Let us take a moment to motivate our model. First off, propagating torsion theories should be taken seriously in their own right. Although typical torsional effects are short-distance, high-energy phenomena, the energy scales are usually considerably lower than those of generic quantum gravity phenomena. Some insight can be gained from the geometric description of elasticity in condensed matter physics. In these solid-state models, defects in an ordered lattice are described using analogous geometric terms like torsion and curvature. There, quantized torsional line-defects (dislocations) are low energy compared to curvature point-defects (disclinations). Thus, it is highly plausible that torsion could begin to propagate in a regime where the spacetime is roughly Minkowski and the curvature of the spin connection is driven by energetics to be approximately zero. This is the regime where our model could apply. 

Secondly, even in strict quantum gravity theories, torsion does play a significant role. For example, many String Theory models predict a propagating Kalb-Ramond field in a low energy effective field theory regime. In gauge theoretic approaches to gravity such as Poincar\'{e} gauge theory and supergravity, the torsion is an intrinsic part of the field. And additionally,  in Loop Quantum Gravity, torsion plays a subtle but fundamental role in the emergence of discrete quantum geometry at the Planck scale as the Barbero-Immirzi term is itself a torsional term. 

Finally, the fermionic features described in our model are likely more generic than might be na\"{i}vely expected since at its root, the fundamental ingredient that allows for the emergence of fermionic geometries in this model is simply a non-linear field theory with a spin structure. 


\section*{The Model}
We will focus our model on the vast arena between the current upper limits of high energy colliders (the TeV scale) and the scale where the backreaction of quantum fields on spacetime curvature  becomes non-negligible. Let us assume that within this regime, torsion not only propagates, but is the dominant geometric degree of freedom. In this regime, spacetime curvature is assumed flat, and the metric can be taken to be Minkowski. Geometry is described by a tetrad $e^I_\mu$, and a spin connection $\omega^{IJ}{}_\nu$. It is convenient to split the spin connection into a Levi-Civita piece together with the contorsion: $\omega^{IJ}{}_\mu=\Gamma^{IJ}{}_\mu+C^{IJ}{}_\mu$. The Levi-Civita connection $\Gamma^{IJ}{}_{\mu}=\Gamma[e]^{IJ}{}_{\mu}$ is the unique tetrad compatible connection satisfying $\p_{[\mu}e^I_{\nu]}+\G^I{}_{J[\mu}e^J_{\nu]}=0$. In terms of the contorsion tensor, the torsion is  $T^I{}_{\mu\nu}\equiv 2\,C^{I}{}_{J[\mu}e^J_{\nu]}$. The spin connection is a $\mathfrak{spin}(3,1)$ valued one-form, and where convenient we will simply write $C_{\mu}$ to denote the $\mathfrak{spin}(3,1)$-valued contorsion one-form. 

Building a model with propagating torsion \cite{hammond2002}, we must to contend with the obvious objection: {\it why hasn't it been measured?} The approach we will take is that torsion has an effective mass. This does two things: (i) a sufficiently large mass gap  prevents the excitation of torsional modes in TeV-scale high-energy collisions and (ii) the torsional field due to an isolated torsional source will take the generic short-range Yukawa form $C\sim \frac{e^{-mr}}{r}.$ The model we propose is that of a {\it constrained \underline{linear} field theory} of massive propagating torsion. Prior to the implementation of the constraints, which we will discuss shortly, we assume each component of the contorsion obeys the Klein-Gordon equation:
\beq
\left(\del_{\mu}\del^\mu -m^2\right) C^{IJ}{}_\nu =0 \label{KG}
\eeq
where $\del_\mu$ is the covariant derivative acting on mixed index tensors that satisfies the tetrad postulate $\del_{\mu}e^I_\nu=0$.\footnote{More specifically, if $\G^I{}_{J\mu}$ is the Levi-Civita connection acting on internal indices that satisfies $\p_{[\mu}e^I_{\nu]}+\G^I{}_{J[\mu}e^J_{\nu]}=0$, and $\G^\alpha{}_{\mu\nu}$ is the ordinary metric compatable Levi-Civita connection satisfying $\p_{\beta}g_{\mu\nu}-\G^\sigma{}_{\mu\beta}g_{\sigma \nu} -\G^{\sigma}{}_{\nu\beta}g_{\mu\sigma}=0$, then the action of $\Del_\mu$ on the mixed index tensor $A^I{}_\nu$ is given by $\Del_\mu A^I{}_\nu=\p_\mu A^I{}_\nu+\G^I{}_{J\mu}A^J{}_\nu-\G^{\sigma}{}_{\nu\mu}A^I{}_\sigma$.}

The physics of the Klein-Gordon equation is well understood and gives rise to free, massive bosonic excitations and thus the interesting results arise when we constrain the theory \cite{Randono:2010cd}. The constraints we will impose are purely geometric, and recall that we are in a regime where back-reaction effects of matter fields are negligible. Thus, the metric, as induced by the tetrad (namely $g_{\mu\nu}=\eta_{IJ}e^I_\mu e^J_\nu$), must be Minkowski, the Levi-Civita connection $\Gamma{}_\mu$ is flat, and $R^{(\Gamma)}_{\mu\nu}=0$. Motivated by the structure of the Einstein-Cartan field equations, we take the lack of backreaction to apply to the curvature of the spin-connection as well. Thus, we impose $R^{(\omega)}_{\mu\nu}=0$. These are our two constraints. In addition, since we are interested in localizable field configurations, we will assume that the torsion approaches zero at asymptotic spatial infinity, thereby allowing for a one-point compactification of a spatial hypersurface:  $\Sigma \cup \{\infty \} \simeq \mathbb{S}^3$.

An action for dynamical torsion with the above field equations and constraints is given by
\beq
S=\frac{\alpha}{2}\int_M \overline{Tr} \left(\,\del_{[\mu} C_{\nu]} \del^\mu C^\nu +\frac{1}{2} m^2 \,C_\mu C^\mu +\lambda_1^{\mu\nu} R^{(\Gamma)}_{\mu\nu} +\lambda_2^{\mu\nu}R^{(\omega)}_{\mu\nu} \right)\, \det(e)\,d^4x
\eeq
\noindent where $\overline{Tr}$ is the trace over the $\mathfrak{spin}(3,1)$ Lie algebra, normalized by the dimension of the representation. The fields $\lambda_1^{\mu\nu}$ and $\lambda_2^{\mu\nu}$ are Lagrange multipliers that implement the constraints. Ignoring these, the field equations of this action reduce to Eq. (\ref{KG}).

The curvature constraints can easily be implemented to isolate the effective degrees of freedom of the constrained theory. First, we note that the flatness of the Levi-Civita and spin connections implies that $\del_{[\mu}C_{\nu]}+C_{[\mu}C_{\nu]}=0$. This does not mean that the contorsion vanishes. In fact,  the constraints imply the existence \cite{Randono:2010cd} of a field $U(x)\in Spin(3,1)$ such that $C_\mu =U \del_\mu U^{-1}$. Thus, the constrained degrees of freedom are entirely contained in the group element $U(x)$. This allows for the construction of a torsional charge $Q$, which is identified with the winding number of the field $U(x)$ viewed as a map $U:\mathbb{S}^3\rightarrow Spin(3,1)$. Since $\pi_3(Spin(3,1))=\mathbb{Z}$, the torsional charge assumes integer values. In contrast to similar constructs in non-Abelian gauge theories, the torsional charge is identically gauge invariant under both large {\it and} small gauge transformations and can thus represent a physical, gauge-invariant charge \cite{Randono:2010cd}.

For convenience, we will adopt the the fundamental unit of length $\ell_0 \equiv \frac{2}{m}$ and set this to unity in the action. The solution to the curvature constraints can then be plugged back into the orginal action to obtain:
\beq
S=\alpha \int_M \overline{Tr} \left(-\del_\mu U \,\del^\mu U^{-1} +\frac{1}{8} \left[ U\del_\mu U^{-1}, U\del_\nu U^{-1}\right] ^2 \right)\, \det (e)\,d^4x\,.
\eeq
This action is a non-linear sigma model with target space $G=Spin(3,1)$. In fact, it is nearly identical in form to a familiar action from pre-Standard Model high energy physics, namely the Skyrme model.\footnote{The apparent difference in sign between the terms in the action from most sources comes from a different choice of signature. In this paper, the signature of the metric is $(-,+,+,+)$.} The Skyrme model was first proposed in the early sixties in order to describe the low energy effective field theory of strongly interacting baryons \cite{Skyrme:1961vq,Skyrme:1962vh}. The model was constructed from the isospin symmetry of pions, and therefore has target space $SU(2)$ as opposed to $Spin(3,1)$. In addition, whereas $SU(2)$ is a purely internal symmetry, independent of spacetime symmetries, in our torsional model the internal $Spin(3,1)$ and external $SO(3,1)$ spacetime symmetries are not independent. Other than that, the two models are identical in form.

\section*{Fermionic Statistics}
As the Skyrme model is an effective field theory for baryons, the analogue of the torsional charge $Q$ is the baryon number, which is proportional to the difference in number between quarks and anti-quarks. It follows that isolated, \emph{even}-baryon number configurations should behave as bosons under exchange whereas \emph{odd}-baryon number configurations can behave as fermions. Indeed, it  is well-known that the non-linear nature of the theory allows for fermionic quantization in the odd baryon number sector \cite{Finkelstein:1968hy,Giulini:1993gd,Williams:Skyrme}. Interestingly, the same analysis applies to our torsional model.

We will now outline the general argument for the existence of  fermionic modes, but adapted to our torsional model. The existence of  Fermi-statistics is a reflection of a topological property of the configuration space, namely the existence of non-contractable loops. We can gain some insight by analogy with the group manifold $SO(3)$. A $2\pi$ rotation in $SO(3)$ can be described by a closed loop in the group manifold. However, this closed loop has the peculiar property that it cannot be deformed by a sequence of continuous transformations to a single point. Specifically, the topological properties of closed loops in the manifold can be characterized by the fundamental group $\pi_1(SO(3))=\mathbb{Z}_2$. Thus, there are two classes of loops, those that are contractable and those that are not. Moreover, traversing along a non-contractable loop twice (e.g. a $2\pi+2\pi=4\pi$ rotation) forms a loop which is contractable. This property allows us to construct the {\it double} cover of the manifold, namely $\overline{SO}(3)=SU(2)$. The non-contractable {\it closed} loops of $SO(3)$ become open curves in $\overline{SO}(3)$ with an end-point identified with $-1$. Thus arise fermionic representations. The same applies to any configuration space $\mathcal{C}$ with the property $\pi_1(\mathcal{C})=\mathbb{Z}_2.$ To construct fermions,  one proceeds by using the universal cover $\overline{\mathcal{C}}$ for quantization. The non-contractable loops in the original configuration space $\mathcal{C}$ are open curves in $\overline{\mathcal{C}}$ that upon quantization are associated with the operator $-1$ acting on states in the Hilbert space. 

We now show that our configuration space has the desired properties. First we note that given any configuration $U_0(\vec{x})$, which we take to be static, any other group field $U(\vec{x},s)$ can be constructed by $U(\vec{x},s)=U_0(\vec{x}) W(\vec{x},s)$ where at each point $W(\vec{x},s)\in Spin(3,1)$. The configuration $U(\vec{x},s)$ describes a parameterized path $\gamma(s)$ in $\mathcal{C}$ as $s$ ranges from $0$ to say, $\pi$. Fixing $W(\vec{x},0)\equiv W_0$ and $W(\vec{x}, \pi)=W_\pi$, where $W_0$ and $W_\pi$ are covariantly constant group fields, describes a {\it closed} loop in configuration space since $C^{IJ}{}_\mu(\vec{x}, s=0)=C^{IJ}{}_\mu(\vec{x},s= \pi)$. At any point in the open interval $s\in (0,\pi)$ the configuration describes a map from $\mathbb{S}^3$ to $Spin(3,1)$ while the two endpoints define constant maps to $Spin(3,1)$. Thus, from a topological perspective, we can take the two endpoints to define the two poles of a four-sphere located at polar angles $0$ and $\pi$. The entire closed loop is then described by a map $W:\mathbb{S}^4\rightarrow Spin(3,1)$. The homotopically distinct maps of this sort are characterized by the homotopy group $\pi_4(Spin(3,1))=\mathbb{Z}_2=\pi_{1}(\mathcal{C}).$  One can therefore replace $\mathcal{C}$ with its double cover $\overline{\mathcal{C}}$ prior to quantization to obtain fermionic states in the Hilbert space.

For the Skyrme model it has been demonstrated \cite{Finkelstein:1968hy, Williams:Skyrme,Giulini:1993gd} that an external {\it or} an internal $2\pi$ rotation of an isolated configuration is a contractable closed loop in $\mathcal{C}$ for even $Q$, and non-contractable for odd $Q$. In addition, the {\it exchange} of two identical isolated configurations is contractable for even $Q$ and non-contractable for odd $Q$. The same analysis applies to our torsional configurations. Thus, upon quantization of $\overline{\mathcal{C}}$, \emph{even} torsional charges must be bosons while  \emph{odd} torsional charges can act as fermions under internal {\it or} external rotations and exchanges. 

\subsection*{Subtleties}
It should be remarked, however, that there are some subtleties. Since the internal and external Lorentz groups are intertwined in the torsional model, one has to be careful about which transformations are closed loops. In particular, an external spacetime Lorentz transformation is generated by the Lie derivative $\mathcal{L}_{\bar{K}}$ where $\bar{K}$ is a Killing vector of the Minkowski metric. On the other hand, the internal Lorentz transformations are generated by the $\mathfrak{spin}(3,1)$ Lie algebra elements $\lambda$ where $\del_\mu \lambda=0$. A true rotation, which in the Hamiltonian framework is generated by the total angular momentum (orbital + spin), is a combination of internal and external rotations generated by $\bar{K}$ and $\lambda$ satisfying\footnote{If a transformation that preserves the form of the metric is an {\it isometry}, one might call a transformation that preserves the form of the tetrad an {\it isotetry!}} $\mathcal{L}_{\bar{K}}e^I_\mu+\lambda^I{}_K e^K_\mu =0$. Now, isolated odd torsional charge configurations $C^{IJ}{}_\mu$, form non-contractable loops in the configuration space under external {\it or} internal $2\pi$ rotations separately. It follows that a $2\pi$ rotation generated by the total angular momentum will be a contractable loop. This is a reflection of the fact that internal and external indices are related to each other via the soldering form $e^\mu_I$. For example, the contorsion tensor defined by $C^{\alpha\beta}{}_\mu\equiv e^\alpha_I e^\beta_J \,C^{IJ}{}_\mu$ behaves as a tensor under external spacetime rotations and a scalar under internal rotations. The converse is true for $C^{IJ}{}_{K}\equiv C^{IJ}{}_\mu e^\mu_K$. Thus a $2\pi$ rotation of these objects is a contractable loop in $\mathcal{C}$ under {\it either} internal {\it or} external rotations. Nevertheless, the mixed index, Lie algebra-valued contorsion one-form $C^{IJ}{}_\mu$ can display the fermionic properties described above.

\subsection*{Geometry shows its face}
Perhaps the most striking difference between our torsional model and the original Skyrme model is that the torsional model is a purely geometric theory. To drive this point home, it is illustrative to work in a different gauge. Recall that the geometry is fixed by specifying the set $\{e^I_\mu, \omega^{JK}{}_\nu\}$, from which the torsion tensor can be extracted. Since the theory is geometric, we are free to make an internal $Spin(3,1)$ gauge transformation and the physics will be the same. In particular, since the spin connection is flat, we can always choose a gauge where $\omega^{JK}{}_\nu=0$. In this gauge, in Cartesian coordinates the tetrad takes the simple form $e^I_\mu=U^I{}_J\,\delta^J_\mu$ for some $U(x)\in Spin(3,1)$. As usual, the torsional charge is the winding number of $U(x)$. A closed path representing a $2\pi$ rotation in this gauge is then a parameterized curve denoted by $\{e^I_\mu(s), \omega^{JK}{}_\nu(s)\}=\{U(x,s)^I{}_J\,\delta^J_\mu, 0\}$. This clearly displays the geometric nature of the torsional Skyrmion -- all of the geometric content of the Skyrmion can be shifted into the tetrad alone. In this sense, the tetrad field itself acquires fermionic properties.

\section*{Concluding Remarks}
We have shown shown that a well-motivated, geometric theory of propagating torsion yields field configurations that can be consistently quantized as fermions. This begs one to question whether the generic procedure of blindly quantizing gravity using bosonic statistics might be a bit na\"{i}ve. Details of the theory aside, the fundamental ingredient allowing for the emergence of fermionic structures in our torsional model is a non-linear field theory with a spin-structure. It is possible, and indeed likely, that some of these features may be retained in full quantum gravity theories. 

Let us close with a forward-looking remark. Though we have chosen to focus on the issue of statistics in dynamical geometry theories, there are potentially many more new and fascinating lines of research that could emerge from models of this sort. Most notably, it is tantalizing that such a simple model of propagating torsion can so closely mimic baryonic matter \cite{MultifacetedSkyrmion,Solitons,TopologicalSolitons}, the statistical properties being just one aspect of this. Potential phenomenological consequences are currently under investigation.

\section*{Acknowledgments}
TLH is supported by National Science Foundation, under grants DMR 0758462 at the University of Illinois, and by ICMT. AR is supported by the NSF International Research Fellowship Grant \#0853116.

\bibliography{TSStatistics}

\end{document}